\documentclass[10pt]{article}
\begin{document}
\title{Almost-periodic time observables for bound quantum systems}
\author{Michael J. W. Hall\\
Theoretical Physics, IAS, \\ Australian National
University,\\
Canberra ACT 0200, Australia}
\date{}
\maketitle



\begin{abstract}
It is shown that a canonical time observable may be defined for any quantum system having a discrete set of energy eigenvalues, thus significantly generalising the known case of time observables for periodic quantum systems (such as the harmonic oscillator). The general case requires the introduction of almost-periodic probability operator measures (POMs), which allow the expectation value of any almost-periodic function to be calculated.  An entropic uncertainty relation for energy and time is obtained which generalises the known uncertainty relation for periodic quantum systems.  While non-periodic quantum systems with discrete energy spectra, such as hydrogen atoms, typically make poor clocks that yield no more than 1 bit of time information,  the anisotropic oscillator provides an interesting exception.  More generally, a canonically conjugate observable may be defined for any Hermitian operator having a discrete spectrum. 
\end{abstract}

\section{Introduction}

The time parameter appearing in the Schr\"{o}dinger equation refers, in practice, to the time shown on a classical clock in the laboratory. However, the physical description of such a clock is expected to be the classical limit of some underlying quantum description. Hence, it is of fundamental interest to consider time observables for quantum systems.

Two basic properties are reasonably expected of a time observable $T$ for a given quantum system. First, the linearity of quantum mechanics requires that the measurement statistics of $T$, for a state described by ket $|\psi\rangle$, should be described by a probability density of the form
\begin{equation} \label{prob}
 p(t|\psi) = \langle \psi | T_t|\psi\rangle . 
\end{equation}
Second, for $T$ to track the evolution time of the system, these statistics should be {\it covariant} under evolution from any initial state $|\psi_0\rangle$ to any final state $|\psi_\tau\rangle$, implying that
\begin{equation} \label{cov} 
p(t|\psi_\tau) = p(t-\tau|\psi_0)  .  
\end{equation}
Thus, for example, if $p(t|\psi_0)$ is initially peaked about $t=0$, then $p(t|\psi_\tau)$ will be peaked about $t=\tau$.

Time observables satisfying (\ref{prob}) and (\ref{cov}) above have been previously defined for the case of a discrete energy spectrum having equally-spaced energy eigenvalues \cite{hel74,hel,hol,pombook}, as well as for the case of a continuous energy spectrum \cite{hol,pombook,kijowski, hol78}.  These cases include, for example, isotropic oscillators and free particles.  However, no analogous time observable has been defined for the case of a Hamiltonian operator $H$ having an {\it arbitrary} discrete energy spectrum.  It is this general discrete case - which includes, for example, bound atomic systems, anisotropic oscillators, and particles bouncing under gravity - that is addressed in this paper.

Note that the operator $T_t$ in equation (\ref{prob}) must be positive, so as to generate positive expectation values for positive functions of $T$. The set $\{T_t\}$ therefore forms a probability operator measure (POM) or positive-operator-valued measure (POVM) \cite{hel,hol,pombook}.  Such POM observables are well known to be essential for describing all possible measurements that may be made on a given quantum system, and may always be represented in terms of measurement of a Hermitian operator on some `apparatus' which has interacted with the system.  A major advantage of the POM formalism is that one does not have to describe such apparatuses and interactions explicitly, when considering the possible measurements which can be made.  In the context of the present paper, the consideration of POM observables is in fact unavoidable -  it is well known that the semi-boundedness of the energy spectrum, for any physical quantum system, implies there can be no (covariant) Hermitian time operator acting on the Hilbert space of the system \cite{hel, hol, pombook} (see also section 6 below). 

As reviewed in section 2, the existence of time observables satisfying equations (\ref{prob}) and (\ref{cov}) is related to finding a measure that satisfies a particular `orthogonality' condition.  It is further shown that the measurement of any such observable is equivalent to first adding noise to the system, and then measuring a particular {\it canonical} time observable, $T^{can}$. This optimality property makes the canonical time observable of especial interest.

In section 3, a canonical time observable is shown to exist for a system with an arbitrary discrete energy spectrum, based on a natural measure for almost-periodic functions.  The corresponding statistics generalise the usual notion of a periodic probability density, and in particular allow the expectation value of any almost-periodic observable to be calculated (i.e., of any function $f(t)$ with a countable Fourier series).  These statistics are closely related to the theory of quantum revivals \cite{robin}, and a semiclassical example is discussed.

Time resolution properties are considered in sections 4 and 5, and it is shown that an `almost-periodic entropy' may be defined for almost-periodic observables, leading to an energy-time entropic uncertainty relation.  Examples, including the hydrogen atom and the harmonic oscillator, indicate that non-periodic systems typically make rather poor `clocks', with the anisotropic oscillator being a notable exception. Finally, some generalisations and measurement aspects are discussed in section 6.

\section{Canonical time observables}

It is shown here that, for any nondegenerate quantum system which is periodic or has a continuous energy spectrum,  there is a corresponding `canonical' time observable $T$ which is optimal over all other possible time observables for the system.  The approach is based on known results in the literature \cite{hel,hol}, but is developed in a manner which may be straightforwardly generalised to the case of systems with arbitrary discrete energy spectra in section 3, and to degenerate energy spectra in section 6.

First, note that the probability density $p(t|\psi)$ in equation (\ref{prob}) acquires its physical meaning via the prediction of expectation values, where the expectation value of an observable function $f(T)$ of any time observable $T$ is given by
\begin{equation} \label{expec}
\langle f(T)\rangle = \mu[\, pf\,] 
\end{equation}
for some suitable measure $\mu$ on the outcome space of $T$ (typically the Lebesgue measure).  
Making the particular choice $f(t)\equiv 1$ yields the normalisation requirement 
\begin{equation} \label{norm}
\mu[\,p\,] = 1 . 
\end{equation} 

To characterise the set of time observables, note that substitution of $t=\tau$ and $|\psi_t\rangle = e^{-iHt/\hbar}|\psi_0\rangle$ into equations (\ref{prob}) and (\ref{cov}) yields the covariance relation $T_t = e^{-iHt/\hbar} T_0 e^{iHt/\hbar}$ \cite{hel,hol}. Inserting this covariance relation into equation (\ref{norm}) and using the completeness of the energy eigenstates then gives
\[  \sum_{E,E'} \langle\psi|E\rangle \langle E'|\psi\rangle \, \langle E|T_0|E'\rangle \, \mu[e^{i(E'-E)t/\hbar}] = 1    = \sum_{E,E'} \langle\psi|E\rangle \langle E'|\psi\rangle \,\langle E|E'\rangle, \]
where $\{|E\rangle\}$ denotes the set of energy eigenstates of the system, and summation is replaced by integration over any continuous parts of the spectrum. Since this must hold for all states $|\psi\rangle$, it follows immediately that $T_0$ must satisfy
\begin{equation} \label{cond}
\langle E|T_0|E'\rangle \, \mu[e^{i(E'-E)t/\hbar}] = \langle E|E'\rangle  
\end{equation}
for all $E$ and $E'$.

To proceed further, it is convenient to assume that the measure $\mu$ satisfies the `orthogonality' property
\begin{equation} \label{orthog}
\mu[\,e^{i(E'-E)t/\hbar}\,] = \gamma \,\langle E|E'\rangle
\end{equation}
for some constant $\gamma >0$ (this assumption will be relaxed slightly in section 6).  This property and equation (\ref{cond}) lead immediately to the fundamental characterisation that $\{T_t\}$ {\it represents a time observable satisfying equations (\ref{prob}) and (\ref{cov}) if and only if}
\begin{equation} \label{t0}
   T_t = e^{-iHt/\hbar} T_0 \,e^{iHt/\hbar},~~~~~T_0\geq 0 ,~~~~~\langle E |T_0|E\rangle = \gamma^{-1} ,
\end{equation}
{\it for all energy eigenvalues $E$}.

Note that the `orthogonality' assumption (\ref{orthog}) is certainly valid for a system with a {\it continuous} energy spectrum: choosing the Lebesgue measure $\mu_c$ over the whole real line, the lefthand side of (\ref{orthog}) is given by
\[ \int_{-\infty}^\infty dt\, e^{i(E'-E)t/\hbar} = 2\pi\hbar \,\delta(E-E')=\gamma_c\, \langle E|E'\rangle,~~~~~\gamma_c:=2\pi\hbar, \]
for this case.  It is also valid for any {\it periodic} system, with period $\tau >0$, where one identifies outcome $t+\tau$ with $t$ and the measure is defined for all periodic functions $g(t)$ by
\[ \mu_\tau[g] = \int_0^\tau dt\, g(t) . \]
In particular, such periodic evolution implies an energy spectrum of the form $\{E_j = e_0 + 2\pi n_j/\tau\}$
for some set of integers $\{ n_j\}$, and hence that
\[ \mu_\tau[e^{i(E - E')t/\hbar}] = \tau \, \delta_{E,E'} =\gamma_\tau\,\langle E|E'\rangle,~~~~~\gamma_\tau:=\tau .  \]
Thus the characterisation of time observables in equation (\ref{t0}) applies to all such systems.

The {\it canonical} time observable, $T^{can}$, is defined by the particular choice \cite{hel,hol}
\begin{equation} \label{can}
T^{can}_0 := \gamma^{-1} \sum_{E,E'} |E\rangle\langle E'|  
\end{equation}
in equation (\ref{t0}). From equations (\ref{prob}) and (\ref{t0}) the corresponding canonical time distribution can be written in the form
\begin{equation} \label{pcan} 
p^{can}(t|\psi) = |\langle t|\psi\rangle|^2, 
\end{equation}
where the (nonorthogonal) `time' kets $|t\rangle$ are defined by the Fourier relation
\begin{equation} \label{tket}
|t\rangle := \gamma^{-1/2}\sum_E e^{-iEt/\hbar} |E\rangle  , 
\end{equation}
analogous to the well known relation between conjugate position and momentum kets.  In this sense the canonical time observable is seen to be conjugate to the energy observable of the system.

It has been previously shown, for the case of quantum systems which are periodic or have a continuous energy spectrum, that the canonical time observable $T^{can}$ is optimal over other time observables for estimating an unknown time shift, relative to various performance measures \cite{hel74,hel, hol, kijowski,hol78,caves}.  However, a more fundamental result can be obtained:  {\it the measurement of any time observable $T$ on a quantum system is equivalent to first subjecting the system to a corresponding `noise' process $\phi_T$, and then measuring the canonical time observable $T^{can}$, i.e.,}
\begin{equation} \label{noise} 
T \equiv T^{can} + {\rm noise} . 
\end{equation} 
Optimality is in this sense independent of any given performance measure, and generalises a similar property for optical phase \cite{hallcan}.

To demonstrate equation (\ref{noise}), note that the positivity of $T_0$ implies one can write $T_0=\sum_m |m\rangle\langle m|$ for some set of kets $\{ |m\rangle\}$.  Now consider the quantum channel defined by the completely positive map
\[ \phi_T(\rho) := \sum_m  A_m\rho A_m^\dagger,~~~~~~~A_m:= \gamma^{1/2} \sum_E |E\rangle\langle E|\,\langle m|E\rangle  \]
for an arbitrary density operator $\rho$.  It is easy to check via (\ref{t0}) that $\sum_m A_m^\dagger A_m =1$, as required for such channels \cite{pombook}.  Further, it follows directly from equations (\ref{prob}), (\ref{t0}) and (\ref{can}) that
\[ p(t|\rho) = {\rm tr}[\rho\,T_t] = {\rm tr}[\phi_T(\rho)\, T^{can}_t] = p^{can}\left(t|\phi_T(\rho)\right).  \]
Thus measurement of $T$ on a given state is equivalent to acting on the state by the process $\phi_T$ and then measuring $T^{can}$, as claimed.  It is also worth noting that the energy statistics of $\phi_T(\rho)$ are identical to those of the original state $\rho$.  Hence, {\it the maximum possible time resolution obtainable under some energy constraint, via measurement of any time observable $T$, is never greater than that obtainable via measurement of $T^{can}$}, generalising an analogous result for optical phase \cite{hallcan,halljmo}.

\section{Almost-periodic time observables}

The characterisation of time observables in equation (\ref{t0}), and the definition of the canonical time observable in equation (\ref{can}), rely only on the `orthogonality' property (\ref{orthog}) assumed  for the measure $\mu$ that generates expectation values as per equation (\ref{expec}). Hence, the results of the previous section can be  immediately generalised to the case of an arbitrary (nondegenerate) discrete energy spectrum, $\{ E_0,E_1,E_2,\dots\}$,  providing that a suitable measure $\mu$ can be defined for this case.  This can indeed be done, as will now be shown.  The corresponding time observables are closely associated with the theory of almost-periodic functions \cite{bohr,bes}.

In particular, following Besicovitch \cite{bes}, consider the {\it almost-periodic }measure
\begin{equation} \label{muap}
\mu_{ap}[g] := \limsup_{X\rightarrow\infty} \frac{1}{X} \int_0^X dt \, g(t) .
\end{equation}
This measure (also called the Besicovitch mean) is well-defined on the algebra of bounded almost-periodic functions $g(t)$, i.e., for functions with a countable Fourier series of the form \cite{bohr,bes}
\begin{equation} \label{gap}	
g(t) = \sum_j a_j\, e^{i\omega_j t},~~~~~~~~\sum_j |a_j|^2 <\infty , 
\end{equation}
including periodic functions in particular. The measure is clearly linear; positive whenever $g$ is positive;  translation-invariant (i.e., the range of integration can be replaced by $[t_0, t_0+X]$ for any $t_0$); and can be shown to satisfy the Parseval relation 
\begin{equation} \label{parseval} 
\mu_{ap}\left[\,|g|^2\,\right] = \sum_j \,|a_j|^2 
\end{equation}
for any bounded almost-periodic function $g$ \cite{bohr, bes}.  

Noting equations (\ref{expec}) and (\ref{norm}), it follows that one may define an {\it almost-periodic probability density} to be any almost-periodic function $p(t)$ satisfying 
\[ p(t) \geq 0,~~~~~~~~~\mu_{ap}[\,p\,] = 1 .  \]
These properties ensure that the expectation values of positive quantities are positive and normalised.  However, for expectation values $\langle f\rangle = \mu_{ap}[\,pf\,]$ to be well-defined with respect to the Besicovitch measure in equation (\ref{muap}), the algebra of observable functions $\{f(t)\}$  must be restricted to the set of almost-periodic functions.  Thus, consistency requires that {\it only almost-periodic observables are described by almost-periodic probability densities}.  This is precisely analogous to the requirement that, for a periodic probability density having period $\tau>0$, one must restrict observables to the algebra of periodic functions of period $\tau$.  This has some interesting implications, as will be seen in section 4.  

The connection with periodic probability densities is worth clarifying a little further here.  In particular, it is straightforward to show that the periodic measure $\mu_\tau$ defined in the previous section satisfies
\[ \mu_\tau[g] = \tau\,\mu_{ap}[g]   \]
for any periodic function $g(t)$ having period $\tau$.  It follows that if $p_\tau(t)$ is a periodic probability density with respect to $\mu_\tau$, then defining a corresponding almost-periodic probablity density by
\begin{equation} \label{pmap}
p_{ap}(t) := \tau\, p_\tau(t) 
\end{equation}
yields
\[ \langle f\rangle = \mu_\tau[\,p_\tau f\,] = \mu[ \,p_{ap}f\,] , \]
for all periodic functions $f(t)$ having period $\tau$. Hence, {\it $\mu_{ap}$ provides an equivalent alternative to $\mu_\tau$ for describing the statistics of periodic observables} (including time observables for periodic quantum systems). 

However, the real advantage of the Besicovitch measure lies in its more general applicability, including to the description of time observables for any quantum system with a discrete energy spectrum, $\{ E_0,E_1,E_2,\dots\}$.  In particular, definition (\ref{muap}) implies that \cite{bohr,bes}
\begin{equation} \label{aporthog}
\mu_{ap}[\,e^{i(E_j-E_k)t/\hbar}\,] = \delta_{jk} = \gamma_{ap}\,\langle E_j|E_k\rangle,~~~~~~\gamma_{ap}:=1 . 
\end{equation}
Hence, the orthogonality property (\ref{orthog}) holds, and it follows immediately from the results of the previous section that the set of time observables, in the nondegenerate case, is characterised by equation (\ref{t0}), with $\gamma=\gamma_{ap}=1$.  Moreover, the canonical time observable $T^{can}$ for the system is defined via equation (\ref{can}), with corresponding almost-periodic probability density given by equations (\ref{pcan}) and (\ref{tket}), and is optimal in the sense discussed with respect to equation (\ref{noise}).  Any POM $\{T_t\}$ generating an almost-periodic probability density via equation (\ref{prob}) may be called an {\it almost-periodic POM}, and satisfies $\mu_{ap}[T_t]=1$. Measurement aspects for such observables are discussed in section 6.

In more detail, if the system at time $\tau$ is described by
\begin{equation} \label{psi} 
|\psi_\tau\rangle = \sum_j c_j e^{-iE_j\tau/\hbar} |E_j\rangle ,~~~~~~\sum_j |c_j|^2=1,   
\end{equation}
then a corresponding `canonical time representation' may be defined by
\begin{equation} \label{theta}
\theta_\tau(t) = \langle t|\psi_\tau\rangle = \sum_j c_je^{iE_j(t-\tau)/\hbar} = \theta_0(t-\tau) ,
\end{equation}
and the canonical time probability density follows via equation (\ref{pcan}) as
\begin{equation} \label{pap}  
p(t|\psi_\tau) = |\theta_\tau(t)|^2 = |\theta_0(t-\tau)|^2 .  
\end{equation}
Comparison of equations (\ref{gap}) and (\ref{theta}) shows that $\theta_\tau(t)$ is an almost-periodic function, as is the related autocorrelation function
\[ A(\tau):=\langle\psi_\tau|\psi_{0}\rangle = \sum_j |c_j|^2 e^{iE_j\tau/\hbar} = \mu_{ap}\left[\, \theta_0^*(t)\,\theta_0(t+\tau)\, \right] , \]
relevant to the description of quantum recurrence and revival times \cite{robin,reviv,nauen}.  Almost-periodic autocorrelation functions have also been studied in classical signal processing contexts, where they are known as almost-periodically correlated processes or almost-cyclostationary processes \cite{classap}.

Known methods for approximating the autocorrelation function $A(\tau)$ \cite{robin,reviv,nauen} may easily be modifed to approximate the time representation $\theta(t)$, since the latter simply corresponds to the replacement of $|c_n|^2$ by $c_n$ in the former.  As an example, consider a semiclassical state described by slowly-varying coefficients $c_n$, that contribute significantly to $|\psi\rangle$ only for relatively large values of $n$ about some average value $\overline{n}$.  Approximating $c_n$ and $E_n$ by slowly-varying continuous functions $f(n-\overline{n})$ and $E(n)$ (note that $E_n\sim n^{2k/(k+2)}$ for a potential $V(x)\sim |x^k|$ \cite{robin}), one may then expand the energy eigenvalues to second order in $n-\overline{n}$ to give the semiclassical approximation
\begin{equation} 
\theta(t) \approx \sum_{n=-\infty}^\infty \,f(n)\, e^{i(n {E}' + n^2{E}''/2)t/\hbar} 
\end{equation}
up to a phase factor, where ${E}'$ and ${E}''$ denote $E'(\overline{n})$ and $E''(\overline{n})$ respectively. This is typically valid up to at least the revival time, $\tau_r=4\pi\hbar/E''$ \cite{robin}, and hence may be used to calculate expectation values of almost-periodic functions $g$ satisfying $2\pi/\omega_j < \tau_r$ for all $j$ in equation (\ref{gap}).  

For the particular case of Gaussian coefficients, with
\[ f(n)  \approx (2\pi\sigma)^{-1/4} \, e^{-n^2/(4\sigma^2)}, ~~~~~\sigma \ll \overline{n}, \]
one may follow Nauenberg \cite{robin,nauen} and apply the Poisson summation formula, to calculate $\theta(t)$ as the sum of relatively displaced Gaussians
\[ \theta(t) \approx (2\pi\sigma)^{-1/4} \sum_{n=-\infty}^\infty [2\pi\alpha(t)]^{-1/2}\, e^{(n-2\pi E't/\hbar)^2/[2\alpha(t)]}
\]
up to a phase factor, with
$\alpha(t)= [(2\sigma^2)^{-1} + 4\pi it/\tau_r]/(4\pi^2)$.  Similarly, for the case of an equally-weighted superposition of $2M+1$ energy eigenstates, i.e.,
\[  f(n) = (2M+1)^{-1/2},~~~~~|n|\leq M \ll \overline{n} , \]
one may again use the Poisson summation formula to approximate the canonical time representation as a sum of relatively displaced Fresnel integrals.

\section{Time resolution and purity}

Equations (\ref{theta}) and (\ref{pap}) yield the explicit form 
\begin{equation} \label{form}
p(t|\psi) = \sum_{jk} c^*_jc_k e^{-(E_j-E_k)t/\hbar} = 1 + \sum_{j\neq k} c^*_jc_k e^{-i(E_j-E_k)t/\hbar} 
\end{equation}
for the canonical time probablity density of state $|\psi\rangle$.  Constructive interference between different contributions to this sum can typically take place at some point only if the corresponding energy differences can be matched in phase.  This is not difficult to arrange for periodic systems, since $2\pi(E_j-E_k)/\hbar$ will always be an integral multiple of the period $\tau$.  However, for generic almost-periodic systems, the ratio of any two energy differences will typically be an irrational number, and so the probability density will not significantly differ from a uniform density.  Hence, almost-periodic quantum systems are not expected to make good clocks in general (although the anisotropic oscillator provides an exception, as will be seen in the following section).

Since expectation values relative to an almost-periodic probablility density $p(t)$ are only well-defined for almost-periodic observables $f(t)$, the moments $\mu_{ap}[p\, t]$ and $\mu_{ap}[p\, t^2]$ are not meaningful as expectation values (and indeed diverge in general).  Hence, the degree to which $p(t)$ is concentrated must be characterised by some quantity other than variance.  Note that the use of variance to characterise time resolution is also problematic for quantum systems having a continuous energy spectrum, as it diverges for any state having a non-zero groundstate energy component \cite{hol}.  Here the purity of $p(t)$ will be investigated as a measure of concentration, and entropy will be considered in section 5.  Only canonical time observables will be considered, in view of their optimal properties as noted in section 2.

The {\it purity} of an almost-periodic probability density is defined to be the quantity
\begin{equation} \label{purity} 
P_{ap} := \mu_{ap}[\,p^2\,] . 
\end{equation}
Since the Schwarz inequality holds for the Besicovitch measure (consider the positive quantity $\mu_{ap}[\,|f-\lambda g|^2\,]$), one has the lower bound
\[ P_{ap} \geq \left( \mu_{ap}[\,p\,]\right)^2 = 1  ,\]
where equality holds only in the limit of a uniform density $p\equiv 1$.  Note from equation (\ref{pmap}) that, for a periodic probability density, one has the relation $P_{ap}=\tau P_\tau$ between almost-periodic and periodic purities, where $P_\tau := \mu_\tau[\,p_\tau^2\,]$.

To illustrate the above point that almost-periodic systems are not typically expected to have good time resolution, note that in the generic case the mapping from $(j,k)$ to $E_j-E_k$ will be one-one for $j\neq k$. It follows in such a case, via (\ref{form}) and the Parseval relation (\ref{parseval}), that the purity is given by
\begin{equation} \label{typical} 
P_{ap} = 1 +\sum_{j\neq k} |c^*_jc_k|^2 = 2 - \sum_{j} |c_j|^4 < 2  ,
\end{equation}
and hence is strictly bounded for such systems.  Note that this result is valid whenever there are no shared resonances between different energy levels, i.e., when there are no distinct pairs of eigenvalues with $E_j-E_k=E_m-E_n$.  This includes, for example, the case of a bound hydrogen atom, where $E_n \sim 1/n^2$, $n=1,2,\dots$.

In contrast, arbitrarily high purities can be obtained for systems with multiple shared resonances.  For example, consider the case of a harmonic oscillator, with energy eigenvalues $E_n=n\hbar\omega$ for $n=0,1,2,\dots$, described by the coherent phase state
\begin{equation} \label{coh}
|u\rangle = (1-u^2)^{1/2}\sum_j u^j \,|E_j\rangle,~~~~~0\leq u<1 . 
\end{equation}
The almost-periodic canonical time probability density may be calculated directly, or from the corresponding periodic phase probability density in equation (37b) of Ref.~\cite{halljmo}, as
\begin{equation} \label{cp} 
p(t) = 1 + \sum_{j=1}^\infty u^j\left( e^{ij\omega t} + e^{-ij\omega t} \right) , 
\end{equation}
and the purity follows directly from equations (\ref{parseval}) and (\ref{purity}) as
\[ P_{ap} = 1 + 2\sum_{j=1}^\infty u^{2j} = \frac{1+u^2}{1-u^2} = 1+2\overline{E}/(\hbar\omega) . \]
Clearly, unlike the typical purity in equation (\ref{typical}), this becomes arbitarily large as the average energy increases. Indeed, coherent phase states are known to have excellent resolution properties with respect to phase \cite{hallcan,halljmo}, and hence, noting $\Phi\sim \omega T$, also with respect to time.

It is of interest to note from equation (\ref{typical}) that, for the `typical' case of an almost-periodic system with no shared resonances, one has the exact uncertainty relation
\begin{equation} \label{exact} \mu_{ap}[ \, p^2\,] + \sum_j (p_j)^2 = 2  
\end{equation}
for the time and energy purities of all pure states, where $\{p_j:=|c_j|^2\}$ denotes the energy distribution.  However, to obtain a time-energy uncertainty relation applicable to {\it all} systems having discrete energy spectra, it is convenient to use a different measure of uncertainty.  This is done in the following section.

\section{Entropic uncertainty relation}

In direct analogy to the continuous and periodic cases, the {\it entropy} of an almost-periodic probability density $p(t)$ is defined to be
\begin{equation} \label{sap}
S_{ap} :=  \mu_{ap}[\,-p\,\log p \, ]  .
\end{equation}
Note from equation (\ref{pmap}) that, for a periodic probability density, this is related to the quantity $S_\tau :=\mu_\tau[\,-p_\tau \log p_\tau\,]$ by $S_{ap}=S_\tau - \log\tau$. Hence entropy differences and relative entropies are invariant with respect to the choice of measure.  Minimising $S_{ap}$ subject to $\mu_{ap}[\, p\, ]=1$ yields the upper bound $S_{ap} \leq 0$, with equality holding only in the limit of a uniform density $p=1$.  It is therefore tempting (and certainly valid for periodic densities \cite{halljmo}) to interpret the quantity 
\begin{equation} \label{iap}
I_{ap}:= -S_{ap} \geq 0  
\end{equation}
as quantifying the information content of $p$, corresponding to the maximum information obtainable from a measurement of $T$ about the value of an unknown time shift applied to the system.

As for purity in the previous section, one can use the entropy to illustrate the poor time resolution expected of typical almost-periodic systems. Indeed, from a direct application of the concavity of the logarithm function one has the inequality
\[ S_{ap} \geq - \log P_{ap} \]
relating entropy and purity.  It follows immediately from equation (\ref{typical}) that $S_{ap}$ is strictly bounded below by $-\log 2$ for the `typical' case of no shared resonances, such as a bound hydrogen atom.
Hence, the corresponding information content $I_{ap}$ is never more than $1$ bit in such cases. In contrast, for the coherent phase state in equation (\ref{coh}), one may use equation (52) of Ref.~\cite{halljmo} to calculate
\[ S_{ap} = \log (1-u^2) = -\log [1+\overline{E}/(\hbar\omega) ] ,  \]
which takes arbitrarily negative values as the average energy increases.

To obtain a more general tradeoff between energy and time resolution, note that the mapping from $\{c_n\}$ to $\theta(t)$ in equation (\ref{theta}) is one-one for a nondegenerate energy spectrum, and satisfies both the Parseval relation (\ref{parseval}) and 
\[ \max |c_j| = \max \left|\mu_{ap}[\,\theta\, e^{-inE_jt/\hbar}\,]\right| \leq \mu_{ap}[\, |\theta|\,] . \]
Hence, the Hausdorff-Young inequality 
\[ (\, \sum_j |c_j|^p \, )^{1/p} \leq \left( \,\mu_{ap}[\,|\theta|^q\,]\,\right)^{1/q},~~~~p^{-1}+q^{-1}=2,~~~~p\geq2, \]
may be obtained via the Riesz-Thorin interpolation theorem in the usual manner \cite{young}, and the method of Bialynicki-Birula and Mycielski \cite{bbm} directly applied to obtain the almost-periodic time-energy entropic uncertainty relation
\begin{equation} \label{eur} 
S(H) + S_{ap}(T) \geq 0 , 
\end{equation}
where $S(H):=-\sum_j|c_j|^2\log |c_j|^2$ denotes the entropy of the energy distribution.  

The entropic uncertainty relation (\ref{eur}) is tight, with equality being achieved for any energy eigenstate, and generalises immediately to mixed states due to the concavity of the entropy.  Note that it may equivalently be expressed as an upper bound for the information content $I_{ap}$ in equation (\ref{iap}):
\begin{equation}  I_{ap}(T) \leq S(H) . 
\end{equation}

Candidates for {\it approximate} minimum uncertainty states, which come close to achieving the lower bound in equation (\ref{eur}), may be generated following the method in section 4 of Ref.~\cite{halljmo} for phase.  In particular, if the maximum possible entropy of the energy distribution under some constraint $C$ is achieved by the probability distribution $\{p_j\}$, then define the state
\[ |C \rangle := \sum_j \sqrt{p_j} \, |E_j\rangle  . \]
This state is a coherent superposition having the maximum possible spread of energy contributions, as measured by $S(H)$, and hence is expected to have a relatively good time resolution.

As an interesting example, consider a two-mode anisotropic oscillator, with Hamiltonian
\[  H=\hbar(\omega_1 a^\dagger a + \omega_2 b^\dagger b),~~~~~[a,a^\dagger]=1=[b,b^\dagger],~~~~~[a,b]=0, \] 
such that the ratio $\omega_1/\omega_2$ is an irrational number.  The energy spectrum is then nondegenerate, with distinct eigenvalues $E_{mn}=m\hbar\omega_1+n\hbar\omega_2$.   Under an average energy constraint, $\langle H\rangle={\cal E}$, the probability distribution maximising $S(H)$ is easily found to be the factorisable (thermodynamic) distribution
\[ p_{mn} = (1-U)(1-V) \,U^{m}V^{n}, ~~~~~U:= e^{-\omega_1/\Omega},~~~~V:=e^{-\omega_2/\Omega},  \]
where $\Omega$ is defined implicitly via
\[ {\cal E} = \hbar \omega_1 U/(1-U) + \hbar \omega_2V/(1-V) . \]
Relatively good time resolution properties are therefore expected for the corresponding state 
\[ |{\cal E}\rangle :=\sum_{mn} (p_{mn})^{1/2}\,|E_{mn}\rangle = |u\rangle\otimes |v\rangle ,\]
where $u=U^{1/2}$, $v=V^{1/2}$, and $|u\rangle$ and $|v\rangle$ are coherent phase states as per equation (\ref{coh}). It is seen that $|{\cal E}\rangle$ is a tensor product of two single-mode oscillators having incommensurate periods $2\pi/\omega_1$ and $2\pi/\omega_2$, and therefore provides a rather simple example of a non-periodic system.  In particular, the use of this state as a `clock' corresponds to using two independent clocks running at incommensurate speeds.

The product form of the state $|{\cal E}\rangle$ allows the time and energy entropies to be calculated as the sum of the corresponding entropies for the two coherent phase states $|u\rangle$ and $|v\rangle$ (where the time entropy for state $|u\rangle$ has been given above), yielding
\[ S(H) + S_{ap}(T) = -[U/(1-U)]\log U - [V/(1-V)]\log V < 2\log e , \]
Comparison with the entropic uncertainty relation (\ref{eur}) confirms that $|{\cal E}\rangle$ is indeed an approximate minimum uncertainty state of time and energy, with the lefthand side being within $2\log e\approx 3$ bits of the minimum possible value.

It is natural to ask what happens in the `isotropic' limit ?  In particular, does a two-mode anisotropic oscillator in the state $|u\rangle\otimes |v\rangle$, with $|\omega_1-\omega_2|\ll \omega:=(\omega_1+\omega_2)/2$, make a better or worse clock than an two-mode isotropic oscillator in the state $|u\rangle\otimes |u\rangle$ with $\omega_1=\omega_2=\omega$ ? 

For the anisotropic case one finds from the above, in the limit of a small (incommensurate) frequency difference, that
\begin{equation} \label{aniso}  
S_{ap}(T_{anisotropic}) \approx 2\log (1-U) \approx -2 \log  \left[ 1+ {\cal E}/(2\hbar\omega)\right].
\end{equation}
The calculation for the isotropic case is a little more complicated, as the energy spectrum is degenerate, with corresponding eigenstates $|E_n,d\rangle=|d\rangle\otimes n-d\rangle$ for $d=0,1,\dots,n$, where $E_n=n\hbar\omega$ and $|j\rangle$ denotes the $j$th energy state of a single-mode oscillator. The canonical time probability density for a degenerate system is defined in equation (\ref{deg}) of section 6, and for the isotropic state $|I\rangle:=|u\rangle\otimes |u\rangle$ under consideration one obtains
\begin{eqnarray*} 
p(t) &=& \sum_{d} |\langle t,d|I\rangle|^2 = \sum_d \left| \sum_{n\geq d}\langle d|u\rangle\langle n-d|u\rangle\, e^{iE_nt/\hbar} \right|^2 \\
&=& (1-u^2)/(1+u^2-2u\cos \omega t) .
\end{eqnarray*}
Remarkably, this is identical to the canonical time probablity density in equation (\ref{cp}) for a single-mode coherent phase state $|u\rangle$, with the two expressions related by a standard trigonometric identity.  It follows immediately that the almost-periodic entropy for the isotropic case is given by
\begin{equation} \label{iso}
S_{ap}(T_{isotropic}) = \log (1-U) = \frac{1}{2} \,S_{ap}(T_{anisotropic})  . 
\end{equation}
Hence, the anisotropic oscillator performs {\it twice} as well as the isotropic oscillator: measuring time via two clocks of slightly different but incommensurate frequencies yields twice as much information than via two clocks of identical frequency, for the above class of states.

Finally, it is worth noting that the anisotropic oscillator remains superior in time resolution when compared to more general states of the isotropic oscillator.  For example, the time entropy for the highly correlated isotropic state
\[ |\chi\rangle := (1-u^2)^{1/2}\sum_m u^m |m\rangle\otimes|m\rangle, ~~~~~{\cal E}=2\hbar\omega u^2/(1-u^2), \]
(which belongs to the nondegenerate subspace of states of the form $\sum_j c_j|j\rangle\otimes |j\rangle$), is again equal to the isotropic entropy in equation (\ref{iso}) above. Further, the time entropy obtained by measuring the {\it single}-mode canonical time observable on the first component of the product $|w\rangle\otimes|0\rangle$ of coherent phase states, with ${\cal E}=\hbar\omega w^2/(1-w^2)$, yields the phase entropy
\[ S_{ap}(T_{single}) = \log (1-w^2) = -\log [1+{\cal E}/(\hbar\omega) ]. \]
This slightly improves on $S_{ap}(T_{isotropic})$ (by up to $\log 2$ as ${\cal E}$ increases), but corresponds to a significantly worse time resolution than the anisotropic entropy in equation (\ref{aniso}) (by $\approx \log [{\cal E}/(4\hbar\omega)]$ for sufficiently large ${\cal E}$). Hence, the anisotropic oscillator provides an exception to the general rule that periodic clocks typically outperform non-periodic clocks.

\section{Discussion}

It has been shown that the existence of time observables satisfying equations (\ref{prob}) and (\ref{cov}) follows from the existence of a suitable measure, $\mu$, satisfying the orthogonality property (\ref{orthog}).  For quantum systems which are periodic or have a continuous energy spectrum, this measure is the Lebesgue measure on an interval or the real line, respectively, and yields the known time observables for these cases.  For a discrete energy spectrum, this measure is the Besicovitch measure in equation (\ref{muap}), and yields almost-periodic time observables. For all cases one can  define canonical time observables, purities, entropies, etc, in a unified manner.

A striking aspect that arises for the case of a discrete energy spectrum is the introduction of almost-periodic probability densities.  These do not appear to have been considered before for quantum systems (although the statistics of almost-periodic functions are of long-standing interest in classical signal processing theory \cite{classap,davis}).  It is therefore worth remarking briefly on some formal and measurability aspects, before generalising results to degenerate energy spectra.

First, it is in fact possible to represent the almost-periodic POM for the canonical time observable as a limit of a sequence of POMs defined with respect to the usual Lebesgue measure. In particular, following a recent suggestion \cite{sign}, for a quantum system with a nondegenerate discrete spectrum define the `normalisation' operator
\[  N(X):= X^{-1}\int_0^X dt \,|t\rangle\langle t|,~~~~~X>0,  \]
where $|t\rangle$ denotes the `time' ket in equation (\ref{tket}) with $\gamma=1$, and let $P_0(X)$ denote the projection operator onto the zero eigenspace of $N(X)$. Equations (\ref{norm}) and (\ref{pcan}) imply that $N(X)\rightarrow 1$, and hence that $P_0(X)\rightarrow 0$, as $X\rightarrow\infty$.  A POM observable $M_X\equiv\{M_t(X);P_0(X)\}$ may then defined for each value of $X$ by
\[  M_t(X):= N(X)^{-1/2}|t\rangle\langle t| N(X)^{-1/2},~~~~~~~t\in [0,X],  \]
where the action of $N(X)^{-1/2}$ is defined to be zero when acting outside the support of $N(X)$, and by construction one has $\int_0^X dt\, M_t +P_0(X) = 1$.  This POM therefore describes a standard quantum observable, and the expectation value of any function $f(t)$ defined on $[0,T]$ follows as
\[ \langle f\rangle_X = X^{-1}\int_0^X  f(t)\left|\langle t|N(X)^{-1/2}|\psi\rangle\right|^2\,dt + \langle \psi|P_0(X)|\psi\rangle  . \]
Taking the limit of the supremum as $X\rightarrow \infty$, it follows that
\begin{equation} \langle f\rangle_\infty = \mu_{ap}[\,pf\, ] \end{equation}
for any almost-periodic function $f$, where $p$ is the almost-periodic probability density of the canonical time observable defined in equation (\ref{pcan}).  

Any almost-periodic POM may similarly be arbitrarily closely represented by a standard POM. Hence, {\it almost-periodic observables are no more or less measurable, in principle, than any other observable}.  It would, of course, be of interest to find a method of measuring an almost-periodic time observable {\it in practice}, at least approximately.  This would complement the approximate measurements of time observables known for harmonic oscillators \cite{wiseman} and free particles \cite{muga}, and, noting section 5, would be of particular interest for anisotropic oscillators. 

It would further be of considerable interest to identify physical measurements on quantum systems that have statistics described by almost-periodic probability densities.  One test for such statistics is that the outcomes $a_1,a_2,\dots$ of a sequence of measurements of an almost-periodic quantity $A$ should have measured expectation values for $e^{-i\omega A}$, i.e., $N^{-1}\sum_{n=1}^N e^{-i\omega a_n}$, that approach zero for all but a discrete set of values $\omega_1$, $\omega_2$, $\dots$, as $N$ increases.   Identifying such a set would further allow the corresponding almost-periodic density $p(a)$ to be approximately reconstructed from the measurement results, via the orthogonality property (\ref{aporthog}), as
\begin{equation} 
p(a)  \approx  \sum_j N^{-1} \sum_{n=1}^N e^{i\omega_j(a-a_n)}  .
\end{equation}
Note that one can also define and investigate almost-periodic {\it discrete} probability densities, $\{ p_1,p_2,\dots\}$, where the expectation value of any function $f(n)$ forming an almost-periodic sequence \cite{cord} is given by 
\[ \langle f\rangle = \mu_d[\,pf\,] = \lim_{N\rightarrow\infty} N^{-1} \sum_{n=1}^N p_n\,f_n . \]

Turning now to the degenerate case, consider first the case of a continuous or discrete energy spectrum which is uniformly degenerate. Thus, the (mutually orthogonal) energy eigenstates have the form $|E,d\rangle$, where the degeneracy index $d$ ranges over some measurable set which is independent of the energy eigenvalue $E$.  It follows that if the generalised orthogonality condition 
\[ \mu[\,e^{i(E'-E)t/\hbar}\,] = \gamma \,\langle E,d|E',d\rangle \]
holds for all $E$, $E'$ and $d$ (which is indeed the case for all measures considered in this paper), then the set of time observables satisfying equations (\ref{prob}) and (\ref{cov}) is characterised by the generalisation 
\begin{equation}   T_t = e^{-iHt/\hbar} T_0 \,e^{iHt/\hbar},~~~~~T_0\geq 0 ,~~~~~\langle E,d|T_0|E,d'\rangle = \gamma^{-1} \delta_{dd'} 
\end{equation}
of equation (\ref{t0}).  Note that this is independent under any relabelling of the degeneracies.  The case of arbitrary degeneracies leads to precisely the same characterisation (and may be obtained by the formal trick of enlarging the Hilbert space to make the energy degeneracy uniform, and then projecting back onto the Hilbert space of physical states).  

The canonical time observable may be defined by labelling the degeneracies for each energy eigenvalue $E$ by $d=1,2,\dots,d(E)$, and choosing
\[  T^{can}_0 = \gamma^{-1} \,\sum_d \sum_{\{E,E':d\leq d(E),d(E')\}} |E,d\rangle\langle E',d| . \]
The results of sections 2-5 then generalise straightforwardly.  In particular, the canonical time probability density for state $|\psi\rangle$ can be written as 
\begin{equation} \label{deg}
p^{can}(t|\psi) = \sum_d |\langle t,d|\psi\rangle|^2,~~~~~|t,d\rangle:= \gamma^{-1/2}\sum_E e^{-iEt/\hbar}|E,d\rangle ,
\end{equation}
where the second summation is over all $E$ such that $d\leq d(E)$ (and is replaced by integration for a continuous energy spectrum).  This expression was used to calculate the time entropy for an isotropic oscillator in equation (\ref{iso}).  

Note that since only the Hermitian property of the Hamiltonian operator has been exploited, these results may also be used to define a canonically conjugate observable for {\it any} Hermitian operator having a continuous or discrete spectrum.  It would be of interest to determine whether such conjugate observables can also be defined for the case of a Hermitian operator having a mixed continuous and discrete energy spectrum.  

Finally, note that the above `time' kets $\{|t,d\rangle\}$ cannot be mutually orthogonal, due to the semiboundedness of the energy spectrum for physical systems.  Hence they cannot be used to define a Hermitian time operator \cite{hel, hol, pombook}. In this respect it is  worth remarking that the `Hermitian time operator' \cite{galapon}
\[ G := i\hbar \sum_{j\neq k} (E_j-E_k)^{-1} |E_j\rangle\langle E_k| \]
recently proposed by Galapon,  for the case of a discrete nondegenerate energy spectrum satisfying $\sum_j (E_j)^{-2}<\infty$, is {\it not} in fact  suitable for defining a time observable. First, it is easily checked that $\langle G\rangle_\tau \neq \langle G\rangle_0 + \tau$ in general, implying that the covariance condition (\ref{cov}) cannot hold.  Thus $G$ does not track the time evolution of the system.  Second, while Galapon claims that the canonical commutation relation $[H,G]=i\hbar$ holds on a dense domain of Hilbert space (note particularly the paragraph following equation (2.18) of \cite{galapon}), one has $[H,G]|E_k\rangle=0$ for any energy eigenstate $|E_k\rangle$, implying this claim is false.  Indeed, the commutation relation holds only on the (noninvariant) subspace $\{\sum_j c_j|E_j\rangle:\sum_j c_j=0\}$, which is manifestly not dense in Hilbert space since for any ket $|\phi\rangle=\sum_j c_j|E_j\rangle$ in this subspace one has 
\[ |\langle\phi|E_k\rangle|^2 = |c_k|^2 = |\sum_{j\,(\neq k)} (-c_j)|^2 \leq \sum_{j\,(\neq k)}|c_j|^2 = \langle\phi|\phi\rangle-|\langle\phi|E_k\rangle|^2 , \]
for any energy eigenstate $|E_k\rangle$, which implies, writing $N=\langle\phi|\phi\rangle$, that 
\[ \left\|\, |\phi\rangle - |E_k\rangle \,\right\|^2 = 1+N - 2\,{\rm Re}\left\{\langle\phi|E_k\rangle\right\} \geq 1+N -(2N)^{1/2} \geq 1/2 . \]
Hence Galapon's operator, while well-defined, has no clear connection with time at all.

\end{document}